\documentstyle[12pt]{article}
\newlength{\extraspace}
\newlength{\extraspaces}

\newcommand{\be}{\begin{equation}}
\newcommand{\ee}{\end{equation}}

\newcommand{\newsection}[1]{
\vspace{15mm}
\pagebreak[3]
\addtocounter{section}{1}
\setcounter{subsection}{0}
\setcounter{footnote}{0}
\setcounter{equation}{0}
\begin{flushleft}
{\large\bf \thesection. #1}
\end{flushleft}
\nopagebreak
\medskip
\nopagebreak}

\newcommand{\ak}{\hat{a}_{k}}
\newcommand{\akp}{\hat{a}_{k'}}
\newcommand{\akd}{\hat{a}^{\dagger}_{k}}

\newcommand{\bk}{\hat{b}_{k}}
\newcommand{\bkp}{\hat{b}_{k'}}
\newcommand{\bkd}{\hat{b}^{\dagger}_{k}}

\begin{document}

\thispagestyle{empty}

\begin{flushright}
{\sc PUPT} 1511, {\sc IASSNS} 94/101\\November 1994
\end{flushright}
\vspace{.3cm}

\begin{center}
{\Large\bf{Effect of Self-Interaction on Charged Black Hole
Radiance}}\\[16mm]
{\sc Per Kraus}\\[3mm]
{\it Joseph Henry Laboratories\\[2mm]
Princeton University\\[2mm]
 Princeton, NJ 08544\\[2mm]
 E-mail: perkraus@puhep1.princeton.edu}
 \\[22mm]

{\sc Frank Wilczek}\\[3mm]
{\it Institute for Advanced Study\\[2mm]
     Princeton, NJ 08540\\[2mm]
     E-mail: wilczek@iassns.bitnet}
 \\[22mm]


{\sc Abstract}

\end{center}

We extend our previous analysis of the modification of the spectrum of
black hole radiance due to the simplest and probably most quantitatively
important back-reaction effect, that is self-gravitational interaction,
to the case of charged holes.  As anticipated, the corrections are
small for low-energy radiation
when the hole is well away from extremality, but  become
qualitatively important near extremality.
A notable result is
that radiation which could
leave the hole with  mass and charge
characteristic of a naked singularity,
predicted in the usual approximation of fixed space-time geometry,
is here suppressed.
We discuss the nature of our approximations, and show
how they work in a simpler electromagnetic analogue problem.

\noindent
\vfill

\newpage

\newsection{Introduction}


Considerable
interest attaches to possible deviations of black hole radiance from
exact thermality.  In a previous paper \cite{KW1}, we showed how inclusion of
the effect of gravitational self-interaction modifies the spectrum,
introducing a definite departure from a thermal distribution.  We did this
by considering the full Hamiltonian for a spherically symmetric,
electrically
neutral black hole interacting with a single particle in the s-wave --
{\it i.e}. a ``shell''.  Upon solving the constraints, we found an
effective particle Hamiltonian.  We analyzed the quantum theory for
this effective Hamiltonian in the WKB approximation, which we
found to be both unambiguous and adequate to describe the late-time
radiation.

In this paper, two additional
things are done.  First, we extend the calculations
to include a charged black hole, and charged matter.  Although this step
does not present any significant formal difficulties,
the physical results we obtain are considerably richer
than what we found in our previous calculations involving neutral holes
and shells.  In the neutral case the final result could be
summarized as a simple replacement of the nominal temperature governing
the radiation by the Hawking temperature for the mass after
radiation, so that the ``Boltzmann factor'' governing emission of energy
$\omega$ from a hole of mass $M$ became
\be
e^{-\omega / T_{\rm eff.} } ~=~ e^{-\omega 8\pi (M -\omega )}~.
\label{teff}
\ee
Note that the argument of the exponential is {\it not\/}
simply proportional to the energy $\omega$, so that the spectrum is
not, strictly speaking, thermal.   While
the deviation from thermality is important in principle its structure,
in this case, is
rather trivial,
and one is left wondering whether that is
a general result.   Fortunately we find that for charged holes the
final results are much more complex.  We say ``fortunately'', not only
because this relieves us of the nagging fear that we have done a
simple calculation in a complicated way, but also for more physical
reasons.
For one knows on general grounds
that the thermal description of black
hole radiance breaks down completely for near-extremal holes \cite{PSSTW}.
One might anticipate, therefore, that something more drastic than a simple
modification of the nominal temperature will occur -- as indeed we
find.  A particularly gratifying consequence of the accurate formula is
a form of ``quantum cosmic censorship''.  Whereas a literal application
of the conventional thermal formulas for radiation yields a non-zero
amplitude for radiation past extremality -- that is, radiation leaving
behind a hole with larger charge than mass -- we find (within our
approximations) {\it vanishing\/} amplitude for such processes.

Second, we discuss in a more detailed fashion the relationship between
our method of calculation, which proceeds by reduction to an effective
particle theory, and more familiar approximations.  We show that it amounts
to saturation of the functional integral of the
underlying s-wave field theory with one-particle intermediate
states, or alternatively to neglect of vacuum polarization.  It is
therefore
closely related to conventional eikonal approximations.  We demonstrate
the reduction of the field theory to a particle theory
explicitly in the related problem of particle creation
by a strong spherically symmetric charge source,
which is a problem of independent
interest.

Before entering the body of this work, for later
reference we here briefly
collect a few formulas describing the Reissner-Nordstrom geometry.
In the rather unconventional gauge \cite{KW2}
whose use we have found to be
extremely convenient, the classical line element for hole of mass
$M$ and charge $Q$  is
\be
ds^2 ~=~ -dt^2 + (dr \pm \sqrt {{2M\over r} - {Q^2\over r^2}} dt)^2
  +  r^2 (d\theta^2 + \sin^2 \theta d\phi^2 )~.
\label{lineelt}
\ee
The outer horizon radius is
\be
R_+ (M, Q) ~=~ M + \sqrt {M^2 - Q^2}
\label{outerhor}
\ee
and the inner horizon radius is
\be
R_- (M, Q) ~=~ M - \sqrt {M^2 - Q^2}~.
\label{innerhor}
\ee
Finally, the nominal Hawking temperature \cite{hawking} is
\be
T (M, Q) ~=~ {1\over 2\pi } { \sqrt {M^2 - Q^2 } \over (M + \sqrt {M^2 -
Q^2})^2 } ~.
\label{nomtemp}
\ee

\newsection{Self-Interaction Correction}

Our system consists of a matter shell of rest mass m and charge q interacting
with the electromagnetic and gravitational fields.  The corresponding action is
\be
S=\int [ -m \sqrt{-\hat{g}_{\mu \nu} d\hat{x}^{\mu} d\hat{x}^{\nu}}
+ q \hat{A}_{\mu} d\hat{x}^{\mu}]
+\frac{1}{16\pi} \int\! d^4\! x \sqrt{-g}\, [{\cal R}-F_{\mu \nu}F^{\mu \nu}]
\label{act}
\ee
$\hat{x}^{\mu}$ is the position of the shell, and a caret over a function
means that it is to be evaluated at the shell ($\hat{f} \equiv
f(\hat{x}^{\mu})$).  To pass to the Hamiltonian formulation we first write
the general spherically symmetric metric in ADM form:
\be
ds^{2} = -N^{t}(t,r)^{2}dt^{2}+L(t,r)^{2}[dr+N^{r}(t,r)dt]^{2}
+R(t,r)^{2}[d\theta^{2}+\sin^{2}\!\theta\, d\phi^{2}].
\label{adm}
\ee
The action can then be written in canonical form as:
\be
S=\int\! dt\,p\,\dot{\hat{r}}~ +~ \int\! dt\, dr\,
[\pi_{R}\dot{R}+\pi_{L}\dot{L}
-N^{t}({\cal H}_{t}^{s} +{\cal H}_{t}^{G} +{\cal H}_{t}^{EM})
-N^{r}({\cal H}_{r}^{s} + {\cal H}_{r}^{G})]~ -~\int\! dt\, M_{\!ADM}
\label{scan}
\ee
with
\be
{\cal H}_{t}^{s}=\left( \sqrt{(p/\hat{L})^{2}+m^{2}}-q\hat{A}_{t}
\right)\delta(r-\hat{r})
\mbox{ \ \ \ ; \ \ \ } {\cal H}_{r}^{s}=-p\,\delta(r-\hat{r})
\label{hshell}
\ee
\be
{\cal H}_{t}^{G}= \frac{L{\pi_{L}}^{2}}{2R^2} - \frac{\pi_{L}\pi_{R}}{R}
+ \left(\frac{RR'}{L}\right)' - \frac{R'^2}{2L} - \frac{L}{2}
\mbox{ \ \ \ ; \ \ \ } {\cal H}_{r}^{G}=R'\pi_{R} - L\pi_{L}'
\label{hgrav}
\ee
\be
{\cal H}_{t}^{EM} = \frac{N^t L {\pi_{\!A_{r}}}^2}{2R^2} - A_t\,
 \pi_{\!A_{r}}'
\label{hem}
\ee
where $'$ represents $d/dr$ and $\dot{}$ represents $d/dt$.  To arrive at this
form we have chosen a gauge such that $A_t$ is the only nonvanishing component
of $A_\mu$.  Of course, we set $A_{r}=0$ only {\em after}
computing the canonical momentum $\pi_{\!A_{r}}$.

Constraints are found by varying the action with respect to $N^t$, $N^r$,
and $A_t$,
$$
{\cal H}_{t} \equiv {\cal H}_{t}^{s}+{\cal H}_{t}^{G}+{\cal H}_{t}^{EM}=0
\mbox{ \ \ \ ; \ \ \ } {\cal H}_{r}\equiv {\cal H}_{r}^{s} +{\cal H}_{r}^{G}=0
$$
\be
\pi_{A_r}'+q\, \delta(r-\hat{r})=0.
\ee
$\pi_R$ can be eliminated by forming the linear combination of constraints
\be
0=\frac{R'}{L}{\cal H}_{t} +\frac{\pi_L}{RL}{\cal H}_{r}
=-{\cal M}' + \frac{R'}{L}({\cal H}_{t}^{s}+{\cal H}_{t}^{EM})
+\frac{\pi_L}{RL}{\cal H}_{r}^{s}
\label{mcon}
\ee
where
\be
{\cal M}=\frac{{\pi_{L}}^2}{2R^2}+\frac{R}{2}-\frac{RR'^2}{2L^2}.
\label{mdef}
\ee
${\cal M}(r)$ is interpreted as being the mass contained within a sphere of
coordinate size r; its value at infinity is the ADM mass.  Similarly,
we see from the Gauss' law constraint that $-\pi_{\!A_r}(r)$ is the charge
contained within a sphere of size $r$, so we define:
$Q(r) \equiv -\pi_{\!A_r}(r)$

Now, if the shell was absent  $ {\cal M}(r) $ and $Q(r)$ would be given by
\be
{\cal M}(r) = M-\int_{r}^{\infty}\! dr\, {R'(r){\cal H}_{t}^{EM}(r)\over L(r)}
\mbox{ \ \ \ ; \ \ \ } Q(r)=Q
\ee
with $M$ and $Q$ being the mass and charge of the black hole as seen from
infinity.  In the gauge $L=1, R=r$ these become
\be
{\cal M}(r)=M-Q^{2}/2r \mbox{ \ \ \ ; \ \ \ } Q(r)=Q.
\label{msol}
\ee
With the shell present we retain the expression (\ref{msol}) for the region
inside the shell, $r<\hat{r}$, whereas outside the shell we write
(with $L=1,\ R=r$),
\be
{\cal M}(r) = M_{+} - (Q+q)^{2}/2r \mbox{ \ \ \ ; \ \ \ } Q(r)=Q+q
\label{out}
\ee
where $M_+$ and $Q+q$ are the mass and charge of the hole-shell system as
measured at infinity.

By using the constraints we can determine $\pi_{R}$, $\pi_{L}$, and an
expression for $M_+$, in terms of the shell variables.  These relations can
then be inserted in the action (\ref{scan}) to give an effective action
depending only on the shell variables.  This program was carried out in
ref. \cite{KW1} for the uncharged case.  The calculation for the present
case runs precisely parallel, resulting in
\be
S=\int\! dt\left[\dot{\hat{r}}\left(\sqrt{2M\hat{r}-Q^{2}} -
\sqrt{2M_{+}\hat{r}
- (Q+q)^{2}\,}\right)
- \eta \dot{\hat{r}}\hat{r} \log{\left|\frac{
\sqrt{\hat{r}}-\eta\sqrt{M_{+}-(Q+q)^{2}/2\hat{r}}}{\sqrt{\hat{r}}-\eta
\sqrt{M-Q^{2}/2\hat{r}}}\right|} - M_{+}\right]
\label{seff}
\ee
where $\eta \equiv$ sgn$\,(p)$, and we have now specialized to a massless shell
($m=0$).  The canonical momentum is then
\be
p_{c} = \sqrt{2M\hat{r}-Q^{2}}-\sqrt{2M_{+} \hat{r}- (Q+q)^{2}}
-\eta \hat{r} \log{\left|\frac{\sqrt{\hat{r}}-\eta \sqrt{M_{+}-(Q+q)^{2}/
2\hat{r}}}{\sqrt{\hat{r}}-\eta\sqrt{M-Q^{2}/2\hat{r}}}\right|}.
\label{pcan}
\ee

At this point we would like to obtain a quantum mechanical wave equation by
making the substitutions $p_{c} \rightarrow -i \partial/\partial r$ ,
$M_{+}-M \rightarrow -i \partial/\partial t$.  However, as discussed in ref.
\cite{KW1}, it is rather difficult to implement this because of the
nonlocal form of (\ref{pcan}).  Fortunately, for our present purposes we need
only compute a class of short wavelength solutions which are accurately
described by the WKB approximation. Writing these solutions as
$v(t,r)=e^{iS(t,r)}$ with $S$ rapidly varying, we can make the replacements
$$
p_{c} \rightarrow \frac{\partial S}{\partial r}
\mbox{ \ \ \ ; \ \ \ } M_{+}-M \rightarrow \frac{\partial S}{\partial t}.
$$
$S(t,r)$ satisfies the Hamilton-Jacobi equation, and so is found by computing
classical action along classical trajectories.  We first choose the initial
conditions for $S(t,r)$ at $t=0$:
\be
S_{k}^{q}(0,r)=kr.
\label{init}
\ee
We have a appended a subscript and a superscript to denote the initial
condition and charge of the solution.   The corresponding classical
trajectory has the initial condition $p_{c}=k$ at $t=0$.  $S_{k}^{q}(t,r)$ is
then given by
\be
S_{k}^{q}(t,r)=k\hat{r}(0) + \int_{\hat{r}(0)}^{r}d\hat{r}\,
 p_{c}(\hat{r}) -(M_{+}-M)t.
\label{sol}
\ee
To determine the radiance from the hole we will will only need to consider
the behaviour of the solutions near the horizon.  Furthermore, only the most
rapidly varying part of the solutions will contribute to the late-time
radiation.  With this in mind, we can write the momentum as (choosing
$\eta =1$ for an outgoing solution)
\be
p_{c}(\hat{r}) \approx -\hat{r} \log{\left| \frac{\hat{r}-R_{+}(M_{+},Q+q)}
{(\hat{r}-R_{+}(M,Q))(\hat{r}-R_{-}(M,Q))}\right|}
\label{pnew}
\ee
so that the initial condition becomes
\be
k=-\hat{r}(0) \log{ \left| \frac{(\hat{r}(0)-R_{+}(M_{+},Q+q))(\hat{r}(0)
-R_{-}(M_{+},Q+q))}{(\hat{r}(0)-R_{+}(M,Q))(\hat{r}(0)-R_{-}(M,Q))}\right|}.
\label{kinit}
\ee
Similarly, the classical trajectory emanating from $\hat{r}(0)$ is given by
approximately,
$$
t \approx \frac{2}{R_{+}(M_{+},Q+q)-R_{-}(M_{+},Q+q)} \left[
R_{+}(M_{+},Q+q)^{2} \log{\left|\frac{\hat{r}-R_{+}(M_{+},Q+q)}{\hat{r}(0)
-R_{+}(M_{+},Q+q)}\right|}\right.
$$
\be
\left. - R_{-}(M_{+},Q+q)^{2} \log{ \left| \frac{
\hat{r}-R_{-}(M_{+},Q+q)}{\hat{r}(0)-R_{-}(M_{+},Q+q)}\right|}\,\right].
\label{geo}
\ee
These trajectories are in fact null geodesics of the metric
\be
ds^2 = -dt^2 + (dr+\sqrt{2M_{+}/r - Q^2}\, dt)^2.
\label{met}
\ee
The relations (\ref{kinit}) and (\ref{geo}) allow us to determine $M_{+}$ and
$\hat{r}(0)$ in terms of the other variables, so that after integrating
(\ref{sol}) we can obtain an expression for $S_{k}^{q}(t,r)$ as a function
of $k$, $t$, and $r$.

We can now write down an expression for the field operator:
\be
\hat{\phi}(t,r)=\int\! dk\, [\hat{a}_{k}v_{k}^{q}(t,r)+\hat{b}_{k}^{\dagger}
v_{k}^{-q}(t,r)^{*}].
\label{phi}
\ee
The modes $v_{k}^{q}(t,r)$ are nonsingular at the horizon, and so the state
of the field is taken to be the vacuum with respect to these modes:
$$
\hat{a}_{k}\left|0_{v}\rangle = \hat{b}_{k} \left|0_{v}\rangle = 0.
\right.\right.
$$
Alternatively, we can consider modes which are positive frequency with
respect to the Killing time $t$.  We write these modes as
$u_{k}^{q}(r) e^{-i\omega_{k}t}$ where the $u_{k}^{q}(r)$ are singular at
the horizon, $r=R_{+}(M+\omega_{k},Q+q)$.  Then
\be
\hat{\phi}(t,r)=\int\! dk\, [\hat{c}_{k}u_{k}^{q}(r) e^{-i\omega_{k}t}
+\hat{d}_{k}^{\dagger}u_{k}^{-q}(r)^{*}e^{i\omega_{k}t}].
\label{phising}
\ee
The two sets of operators are related by Bogoliubov coefficients,
\be
\hat{c}_{k} = \int\! dk\, [\alpha_{kk'} \hat{a}_{k'} +\beta_{kk'}\hat{b}_{k'}
^{\dagger}].
\ee
The flux of outgoing particles of charge $q$ with energy between $\omega_{k}$
and $\omega_{k} + d\omega_{k}$ is given by
\be
F(\omega_{k})=\frac{d\omega_{k}}{2\pi} \frac{\Gamma(\omega_{k})}
{|\alpha_{kk'}/\beta_{kk'}|^{2} - 1}
\ee
where $\Gamma(\omega_{k})$ is a grey-body factor.  This identifies
$|\beta_{kk'}/\alpha_{kk'}|^{2}$ as the effective Boltzmann factor.  From
(\ref{phi},\ \ref{phising}) $\alpha_{kk'}$ and $\beta_{kk'}$ are found to be
$$
\alpha_{kk'} = \frac{1}{2\pi u_{k}^{q}(r)}\int_{-\infty}^{\infty}\! dt\,
e^{i\omega_{k}t}v_{k'}^{q}(t,r)
$$
\be
\beta_{kk'}=\frac{1}{2\pi u_{k}^{q}(r)}\int_{-\infty}^{\infty}\! dt\,
e^{i\omega_{k}t}v_{k'}^{-q}(t,r)^{*}.
\label{bog}
\ee
Here, $r$ is taken to be slightly outside the horizon,
$r=R_{+}(M+\omega_{k},Q+q) + \epsilon$.  These coefficients can be evaluated
in the saddle point approximation.  Recalling that $v_{k}^{q}(t,r)
=e^{iS_{k}^{q}(t,r)}$, the saddle point equation for $\alpha_{kk'}$ becomes
\be
\omega_{k}=-\frac{\partial S_{k'}^{q}}{\partial t} = M_{+}^{q} - M.
\label{alpha}
\ee
This leads to a purely real value of $t$ for the saddle point.
For $\beta_{kk'}$ we have
\be
\omega_{k}=\frac{\partial S_{k'}^{-q}}{\partial t}=M-M_{+}^{-q}.
\label{beta}
\ee
\mbox{}From (\ref{kinit},\ \ref{geo}) we find that the saddle point
value for t has an
imaginary part given by
\be
\mbox{Im}(t_{s})=\frac{2\,R_{+}(M-\omega_{k},Q-q)^{2}}{R_{+}(M-\omega_{k},Q-q)
-R_{-}(M-\omega_{k},Q-q)}\,\pi=\frac{1}{2\,T(M-\omega_{k},Q-q)}.
\ee
Therefore,
\be
|{\beta_{kk'}/\alpha_{kk'}}|
= \frac{1}{|2\pi u_{k}(r)|}\exp{\left(\omega_{k}/T(M-
\omega_{k},Q-q) + \mbox{Im}[S_{k'}^{-q}(t_{s})^{*}]\right)}.
\ee
The terms in $S_{k'}^{-q}$ which contribute to the second term in the
exponent are
$$
\int_{\hat{r}(0)}^{r} d\hat{r}\, p_{c}(\hat{r}) +\omega_{k}\,\mbox{Im}(t_{s}).
$$
Using (\ref{pnew}-\ref{geo}) this can be evaluated to give
\be
\mbox{Im}[S_{k'}^{-q}(t_{s})^{*}]=\frac{M\omega + \sqrt{M^{2}-Q^{2}\,}
\left(\sqrt{(M-\omega)^{2}-(Q-q)^{2}}-\sqrt{M^{2}-Q^{2}}\right)}
{2\,T(M-\omega,Q-q)\, R_{+}(M,Q)}
\ee
resulting in
\be
\left|\frac{\beta_{kk'}}{\alpha_{kk'}}\right|^{2}
=\exp{\left(-\frac{\sqrt{M^{2}-Q^{2}}\, [\omega - \sqrt{(M-\omega)^{2}
-(Q-q)^{2}}+\sqrt{M^{2}-Q^{2}\,}\,]}{T(M-\omega,Q-q)\,R_{+}(M,Q)}\right)}.
\label{bol}
\ee
This is the effective Boltzmann factor governing emission.
Sufficiently far from extremality, when $\omega$, $q \ll
\sqrt{M^{2}-Q^{2}}$, we can expand (\ref{bol}) to give
\be
\left|\frac{\beta_{kk'}}{\alpha_{kk'}}\right|^{2} \approx \exp{\left(-\frac{
\omega - \frac{Qq}{R_{+}(M,Q)} + \frac{M^{2}q^{2}+Q^{2}\omega^{2}
-2MQ\omega q}
{2(M^{2}-Q^{2})R_{+}(M,Q)}}{T(M-\omega,Q-q)}\right)}
\ee
as compared to the free field theory result \cite{hawking},
\be
\left|\frac{\beta_{kk'}}{\alpha_{kk'}}\right|^{2} =
\exp{\left(-\frac{\omega-\frac{Qq}{R_{+}(M,Q)}}{T(M,Q)}\right)}.
\label{free}
\ee
Near extremality, the self-interaction corrections cause the emission to
differ substantially from (\ref{free}).

We might ask whether it is possible to reach extremality after a finite
number of emissions.  Since $T(M-\omega,Q-q)$ appears in the denominator
of the exponent of (\ref{bol}), the transition probability to the extremal
state is in fact zero.  We can also ask whether there are transitions to a
meta-extremal ($Q>M$) hole.  This would have rather dramatic implications
as the meta-extremal hole is a naked singularity.  To address this question
we return to the saddle point equation (\ref{beta}). When $Q>M$, $R_{+}$
and $R_{-}$ become complex.  From (\ref{kinit}) we see that a saddle point
solution would require that $k$ be complex, but we do not allow this since a
complete family of initial conditions $S_{k}(0,r)=kr$ was defined with $k$
real.  Therefore, in the saddle point approximation the extremal hole is
stable.

Modes with $|\beta /\alpha | > 1$ formally require larger amplitudes
for higher occupation numbers, and thus require special interpretation.
Considering for simplicity the free field form of these
coefficients, (\ref{free}), we see that such modes occur when
$\omega < q Q/ R_+$, that is when the incremental energy gain from
discharging the Coulomb field overbalances the cost of creating
the charged particle.  Under these conditions one has dielectric
breakdown of the vacuum, just as for a uniform electric field in
empty space.  Since this physics is not our primary concern in the
present note, we shall restrict ourselves to a few remarks.
The occupation factor appearing in the formula for
radiation in these ``superradiant''
modes is negative, but the reflection probability
exceeds unity, so the radiation flux is positive as it should be.
And in general the formulas for physical quantities will appear sensible,
although Fock space occupation
numbers are not.  We can avoid superradiance altogether by considering
a model with only
massive charged fundamental particles, and holes with a charge/mass ratio
small compared to the minimal value for fundamental quanta.

Another interesting variant is to consider a {\it magnetically\/} charged
hole interacting with neutral matter.  In that case, one simply
puts $q=0$ in the formulae above (but $Q \ne 0$).  One could also consider
the interaction of dyonic holes with charged matter, and other variants
({\it e.g}. dilaton black holes) but we shall not do that here.

\newsection{Discussion}

We have arrived at the result (\ref{bol}) by what may appear to be a
somewhat circuitous route.  Inspired by a field theory question, we
calculated the solutions of a single self-gravitating particle at the
horizon, and then passed back to field theory by interpreting the solutions
as the modes of a second quantized field operator.  In this section we hope
to clarify the logic of this procedure, and show that it is both correct and
efficient, by demonstrating how a single particle action emerges from the
truncation of a complete field theory.

We can illustrate this explicitly if we consider the simpler model of
spherically symmetric
electromagnetic and charged scalar fields interacting in flat space.  Our
goal is to show that the propagator for the scalar field can be expressed as a
Hamiltonian path integral for a single charged shell.
To achieve this, two important
approximations will be made.  The first is that the effects of vacuum
polarization will be assumed to be small, so we can  ignore scalar loop
diagrams.  The second is to assume that the dominant interactions involve
soft photons, so that the difference in the scalar particle's energy before
and after emission or absorption of a photon is small compared to the energy
itself.   Thus we expect that our expression will be valid for cases where
the scalar particle has a large energy, so that the energy transfer per
photon is relatively small, and is far from the origin, so that
the classical electromagnetic self energy of the particle is a slowly varying
function of the radial coordinate.  Field theory in this domain is in fact
well described by the eikonal approximation, which implements the same
approximations we have just outlined.  What follows is then essentially a
Hamiltonian version of the eikonal method.

We start from the action
$$
S=-\frac{1}{4\pi}\int \!d^{4}x\,\left[(\partial_{\mu}-iqA_{\mu})\phi^{*}
\, (\partial^{\mu}+iqA^{\mu})\phi +m^{2}\phi^{*}\phi
+\frac{1}{4}F_{\mu \nu}F^{\mu \nu}\right]
$$
\be
=\int\!dt\,dr\left[\pi_{\phi^{*}}\dot{\phi}^{*}+\pi_{\phi}\dot{\phi}
-\left(\frac{\pi_{\phi^{*}}\pi_{\phi}}{r^2}+r^{2}{\phi^{*}}'
\phi'+m^{2}r^{2}\phi^{*}\phi+\frac{{\pi_{\!A_{r}}}^{2}}{2r^2}
\right)-A_{t}\left(iq[\pi_{\phi^{*}}\phi^{*}-\pi_{\phi}\phi]-\pi_{\!A_{r}}'
\right)\right].
\ee
Defining the charge density
\be
\rho(r)\equiv iq[\pi_{\phi^{*}}(r)\phi^{*}(r)-\pi_{\phi}(r)\phi(r)]
\ee
the solution of the Gauss' law constraint is
\be
Q(r)\equiv -\pi_{\!A_{r}}=\int_{0}^{r}\!dr\,\rho(r)
\ee
and so the scalar field Hamiltonian becomes
\be
H=\int_{0}^{\infty}\!dr\left[
\frac{\pi_{\phi^{*}}\pi_{\phi}}{r^2}+r^{2}{\phi^{*}}'\phi'+m^{2}\phi^{*}\phi
+\frac{Q(r)^2}{2r^2}\right].
\ee
The fields are now written as second quantized operators:
$$
\hat{\phi}=\int\!\frac{dk}{\sqrt{2\pi\,2\omega_k}}\,
\frac{[\ak e^{ikr}+\bkd e^{-ikr}]}{r}
$$
\be
\hat{\pi}_{\phi}=i\int\!\frac{dk}{\sqrt{2\pi}}\sqrt{\frac{\omega_k}{2}}\,r\,
[\akd e^{-ikr}-\bk e^{ikr}]
\ee
where $\omega_{k}=\sqrt{k^{2}+m^{2}}$, and we also have $\hat{\phi}^{*}
=\hat{\phi}^{\dagger}$ , $\hat{\pi}_{\phi^{*}}=\hat{\pi_{\phi}}^{\dagger}$.  To
ensure that the field is nonsingular at the origin we impose the conditions
$\hat{a}_{-k}=-\ak$ , $\hat{b}_{-k}=-\bk$, and take the limits of all $k$
integrals to be from $-\infty$ to $\infty$.

We now write the Hamiltonian in terms of the creation and annihilation
operators.  In doing so we shall normal order the operators, which corresponds
to omitting vacuum polarization since we do not allow particle-antiparticle
pairs to be created out of the vacuum.
 Also when evaluating $\phi'$ we shall use the geometrical optics
approximation, $(e^{ikr}/r)'\approx ike^{ikr}/r$, valid for $k\gg 1/r$.
Then the quadratic part of the Hamiltonian becomes,
\be
\int_{0}^{\infty}\!dr\left[\frac{\hat{\pi}_{\phi^{*}}\hat{\pi}_{\phi}}{r^2}
+r^{2}  \hat{\phi}^{*}{'} \hat{\phi}'+m^{2}\hat{\phi}^{*}\hat{\phi}\right]
=\frac{1}{2}\int\!dk\,\omega_{k}[\akd\ak +\bkd\bk].
\ee
Next we consider the interaction term.  When evaluating this there will arise
factors of $\sqrt{\omega_{k'}/\omega_{k}}$.  The essence of the soft photon
approximation is that we replace these factors by $1$, since we are assuming
that $\Delta \omega/\omega \ll 1$ for the emission or absorption of a single
photon.  Then, after normal ordering, we can evaluate the charge density to
be:
\be
\hat{\rho}(r)=q\int\!\frac{dk\,dk'}{2\pi}[\akd\akp-\bkd\bkp]e^{i(k-k')r}.
\ee

We now wish to calculate matrix elements of the Hamiltonian between one
particle states. A basis of one particle states labelled by position is given
by
\be
|r\rangle=\int\!\!\frac{dk}{\sqrt{2\pi}}\,e^{-ikr}\,\akd|0\rangle.
\ee
The free part of the Hamiltonian then has matrix elements
\be
\langle r_{2}|\,\frac{1}{2}\int\!dk\,\omega_{k}[\akd\ak+\bkd\bk]|r_{1}\rangle
=\int\!\frac{dk}{2\pi}\,\omega_{k}[e^{ik(r_{2}-r_{1})}-e^{ik(r_{2}+r_{1})}].
\ee
The second term in the brackets corresponds to the path from $r_1$ to $r_2$
which passes through the origin.  These paths will not contribute to local
processes far from the origin, so we drop this term.  The matrix elements
of the interaction term for closely spaced points $r_1$ and $r_2$ are:
\be
\langle r_{2}|\int_{0}^{\infty}\!\!dr\,\frac{\hat{Q}(r)^{2}}{2r^2}|r_{1}\rangle
=\frac{q^{2}}{2r_{1}}\int\!\frac{dk}{2\pi}e^{ik(r_{2}-r_{1})} .
\ee
Putting these together, we find the matrix elements of the Hamiltonian,

\be
\langle r_{2}|\hat{H}|r_{1}\rangle=
\int\! \frac{dk}{2\pi}\, e^{ik(r_{2}-r_{1})}
(\sqrt{k^{2}+m^{2}}+q^{2}/2 r_{1}).
\ee
Now we can follow the standard route which leads from matrix elements of
the Hamiltonian to a path integral expression for the time evolution
operator, with the result
\be
\langle r_{f}|e^{-i\hat{H}t}|r_{i}\rangle
=\int_{r(0)=r_{i}}^{r(t)=r_{f}}{\cal D}p\,{\cal D}r\,
e^{i\int_{0}^{t}\!dt'\,(p\dot{r}-\sqrt{p^{2}+m^{2}}-q^{2}/2r)}.
\ee
The action in the exponent is precisely that of a charged shell, with
$q^{2}/2r$ being the electromagnetic self energy.

We now discuss how this analysis can be applied to the case where
we include gravitational interactions.  The resulting field Hamiltonian
is much more complex, and so we will not be able to explicitly calculate the
effective shell action.  However, the preceding derivation allows us to argue
that were we to do so, we would simply derive the effective action obtained in
section 2. The nature of the black hole radiance calculation makes us believe
that the approximations used to arrive at a shell action are justified.  This
is so because for a large ($M\gg M_{\mbox{pl}}$) hole the relevant field
configurations are short wavelength solutions moving in a region of
relatively low curvature, and these are the conditions which we argued make
the eikonal approximation valid.

For simplicity, we will consider an uncharged self-gravitating scalar field.
If we truncate to the s-wave we arrive at what is known as the BCMN model,
originally considered in \cite{BCMN} and corrected in \cite{unruh}. The
action is
$$
S=\frac{1}{4\pi}\int\! d^4\!x\, \sqrt{-g}\left[\frac{1}{4}{\cal R}
- \frac{1}{2}g^{\mu \nu}\partial_{\mu}\phi\partial_{\nu}\phi\right]
$$
\be
=\int\! dt\, dr\,\left[\pi_{\phi}\dot{\phi}+\pi_{R}\dot{R}+\pi_{L}\dot{L}
-N^{t}({\cal H}_{t}^{\phi}+{\cal H}_{t}^{G})-N^{r}({\cal H}_{r}^{\phi}
+{\cal H}_{r}^{G})\right] -\int\! dt\, M_{ADM}
\label{sphi}
\ee
with
\be
{\cal H}_{t}^{\phi}=\frac{1}{2}\left(\frac{{\pi_{\phi}}^{2}}{LR^2}
+\frac{R^2}{L}{\phi '}^{2}\right) \mbox{ \ \ \ ; \ \ \ }
{\cal H}_{r}^{\phi}=\pi_{\phi}\phi'.
\ee
The analog of (\ref{mcon}) is now
\be
{\cal M}'=\frac{R'}{L}{\cal H}_{t}^{\phi}+\frac{\pi_{L}}{RL}{\cal H}_{r}^{s}
=\frac{R'}{2L^2}\left(\frac{{\pi_{\phi}}^{2}}{R^2}+R^{2}{\phi'}^{2}\right)
+\frac{\pi_{L}\pi_{\phi}\phi'}{RL}
\label{mphi}
\ee
The Hamiltonian is
\be
H=M_{ADM}={\cal M}(\infty)=M+\int_{0}^{\infty}\!dr\left[\frac{R'}{L}
{\cal H}_{t}^{\phi}+\frac{\pi_{L}}{RL}{\cal H}_{r}^{s}\right].
\ee

To obtain an expression for $H$ which depends only on $\phi$ and $\pi_{\phi}$
we must choose a gauge and solve the constraints.  We can obtain an explicit
result if we choose the gauge $R=r$, $\pi_{L}=0$.  Then, defining
\be
 h(r)\equiv
\frac{1}{2}\left(\frac{{\pi_{\phi}}^{2}}{r^2}+r^{2}\phi'^{2}\right),
\ee
$L$ is determined from (\ref{mphi}),
\be
{\cal M}'(r)=\left(\frac{r}{2}-\frac{r}{2L^2}\right)'=\frac{h(r)}{L^2}
\ee
so
\be
\frac{1}{L^2}=-\frac{2M}{r}e^{-2\int_{0}^{r}\!dr'\,h(r')/r'}
+\frac{1}{r}e^{-2\int_{0}^{r}\!dr'\,h(r')/r'}
\int_{0}^{r}\!dr'\,e^{2\int_{0}^{r'}\!dr''\,h(r'')/r''}
\ee
which then leads to
\be
H=Me^{-2\int_{0}^{\infty}\!dr\, h(r)/r}
+\int_{0}^{\infty}\!dr\,h(r)e^{-2\int_{r}^{\infty}\!dr'\,
h(r')/r'}.
\label{phiham}
\ee
This generalizes the result of \cite{unruh} to include a nonzero mass $M$
for the pure gravity solution.  To make a direct comparison with our work in
the previous section, it would be preferable to obtain the Hamiltonian in
$L=1, R=r$ gauge.  This is more difficult and we do not know the explicit
expression.   For the moment, though, we are mainly interested in the
qualitative structure of the Hamiltonian, and (\ref{phiham}) will be
sufficient for our purposes.  The various nonlocal
terms contained in the Hamiltonian (\ref{phiham}) correspond to gravitons
attaching onto the particle's worldline.
If we expand the exponentials in (\ref{phiham}), we see that
there arise an infinite series of bi-local, tri-local,  \ldots, terms resulting
from the non-linearity of gravity. Now we could, in principle, repeat the
analysis which led to an effective shell action  for the charged field in
flat space.   In that case the calculation could be done with only modest
effort because there was only a single quartic interaction term. In the
present case we would have to sum the infinite series of terms that arise;
our point is that handling all of these terms is cumbersome,
to say the least, and that it is much simpler to proceed as in
section 2.

\end{document}